\documentstyle[pra,aps,multicol]{revtex}

\input{epsf}

\begin{document}
\draft

\title{Stochastic wave function method for non-Markovian quantum
master equations}

\author{Heinz--Peter Breuer, Bernd Kappler and Francesco Petruccione}
\address{
Albert-Ludwigs-Universit\"at, Fakult\"at f\"ur Physik, \\
Hermann-Herder Stra{\ss}e 3, D--79104 Freiburg im Breisgau,
Federal Republic of Germany}
\date{\today}
\maketitle

\begin{abstract}
A generalization of the stochastic wave function method to quantum
master equations which are not in Lindblad form is developed. The
proposed stochastic unravelling is based on a description of the
reduced system in a doubled Hilbert space and it is shown, that this
method is capable of simulating quantum master equations with negative
transition rates. Non-Markovian effects in the reduced systems
dynamics can be treated within this approach by employing the
time-convolutionless projection operator technique. This ansatz yields
a systematic perturbative expansion of the reduced systems dynamics in
the coupling strength.  Several examples such as the damped Jaynes
Cummings model and the spontaneous decay of a two-level system into a
photonic band gap are discussed. The power as well as the limitations
of the method are demonstrated.
\end{abstract}

\begin{multicols}{2}
\narrowtext

\section{Introduction}
\label{sec:Intro}

The theory of open quantum systems is a fundamental approach to the
understanding of dissipation on a microscopic and macroscopic level in
many fields of physics, such as quantum optics, and solid state
physics. Besides the conventional density matrix formalism, there has
been an increasing interest over the last years in the stochastic wave
function method
\cite{Carmichael,MolmerPRL68,GardinerPRA46,ZollerPRA46,Gisin:92,Gisin:93},
where the state of the open system is described by an ensemble of pure
states instead of a reduced density matrix.  This method permits a
description of the dynamics of an individual quantum system subject to
a continuous measurement \cite{WisemanPRA93,BP:QS8} and hence
provides additional information about the state of the system compared
to the description by a reduced density matrix. Moreover, the
stochastic wave functions method has been shown to be an effective
numerical tool for the solution of density matrix equations with many
degrees of freedom since a reduced density matrix has $N^2$ degrees
of freedom, whereas a stochastic wave vector only has $N$ components
\cite{MolmerPRL74,BP:QS11}.  

Since the stochastic wave function method
originated in the context of quantum optics most publications on this
subject consider the weak coupling regime and the Born-Markov
approximation seems to be inevitably connected with this approach. On
the other hand, the stochastic wave functions method could also
provide a useful numerical tool in other fields of physics, where the
Born-Markov approximation is not justified. This has been shown for
example by Imamoglu \cite{Imamoglu:PRA50}, who extended the stochastic
wave function method to the strong coupling regime by considering an
enlarged system which contains a few ficticious bath-modes, that are
weakly coupled to a Markovian environment.  Although this concept is
quite general, the drawback of this method is obvious: even if the
state space of the system is small, the numerical treatment of the
enlarged system can become very expensive if more than a few
ficticious modes are needed to approximate the reduced systems
dynamics. Another approach to a non--Markovian stochastic wave function 
method developed by Jack, Collett and Walls \cite{WallsNM} is based on a
continuous measurement interpretation of the stochastic unraveling. 
In this approach the stochastic equation of motion for the reduced state 
vector involves a multiple time integration over the system's history
conditioned on the measurement record over a finite time interval.
Furthermore, it has been claimed recently by Di{\'o}si et
al. \cite{DiosiNM97,DiosiNM98} that it is in principle possible to
construct an exact stochastic Schr\"odinger equation which describes
the non-Markovian time evolution of an open quantum system.

In this article we present an extension of the stochastic wave
function method beyond the Born-Markov approximation which is based on
the time-convolutionless projection operator technique
\cite{ShibataZPhysB,ShibataJStat}. This ansatz yields a systematic
perturbative expansion scheme for the stochastic dynamics of the
reduced system which is valid in an intermediate coupling regime where
non-Markovian effects are important, but a perturbative expansion is
still justified. The major advantage of our method is that it does not
rely on an enlarged phase space and that it uses a stochastic
evolution equation which is local in time.

The paper is organized as follows. In Sec.~\ref{sec:GME} we briefly
review three different approaches to equations of motion for the reduced 
density matrix: the derivation of
the Markovian quantum master equation (Sec.~\ref{sec:QME}), the
Nakajima-Zwanzig projection operator technique (Sec.~\ref{sec:Naka}),
which yields a generalized master equation, 
and the time-convolutionless projection operator technique
(Sec.~\ref{sec:tcl}), leading to a quantum master equation which is
local in time. Sec.~\ref{sec:stqme} deals with the stochastic
unravelling of quantum master equations: in Sec.~\ref{sec:stlind} we
review the stochastic unravelling of quantum master equations which
are in Lindblad form, such as the Markovian quantum master equation,
whereas in Sec.~\ref{sec:SGQME} we present an unravelling of arbitrary
linear density matrix equations which are local in time, such as the
time-convolutionless quantum master equation. These algorithms are then
applied to the damped Jaynes-Cummings model and the spontaneous decay
of a two-level system into a photonic band gap in Sec.~\ref{sec:two-lev}.
Sec. V contains our summary.

\section{Derivations of quantum master equations}
\label{sec:GME}

We shall begin with a description of the models we want to examine
and state the basic assumptions underlying the following sections.
Throughout this article, we consider a quantum mechanical system S which is
coupled to a reservoir R. The combined system is supposed to be
closed and its Hamiltonian is of the form
\begin{equation}
  \label{eq:H_tot}
  H=H_{0}+\alpha H_{\rm I},
\end{equation}
where $H_0$ is the Hamiltonian of the system and the reservoir and
$H_{\rm I}$ the interaction Hamiltonian. The time evolution of the
combined system's density matrix in the interaction picture $W(t)$ is
determined by the Liouville-von Neumann equation
\begin{equation}
  \label{eq:LvN}
  \frac{\partial}{\partial t}W(t)=-i\alpha[H_{\rm I}(t),W(t)]\equiv
  \alpha L(t)W(t),
\end{equation}
where the interaction Hamiltonian in the interaction picture is
defined as $H_{\rm I}(t) = \exp(iH_{0}t) H_{\rm I} \exp(-iH_{0}t)$.
The initial state of the combined system is supposed to factorize,
\begin{equation}
  \label{eq:ini}
  W(0)=\rho(0)\otimes\rho_{\rm R},
\end{equation}
where $\rho_{\rm R}$ is some stationary state of the reservoir, i.~e.,
the system and the reservoir are initially uncorrelated. For technical
simplicity, we further assume that odd moments of $H_{\rm I}(t)$ with
respect to $\rho_{\rm R}$ vanish, i.~e.,
\begin{equation}
  \label{eq:odd}
  \mbox{Tr}_{\rm R}\left\{\rho_{\rm R}H_{\rm I}(t_1)\cdots H_{\rm I}(t_{2k+1})\right\}=0,
\end{equation}
although this assumption is not essential for the methods we want to use in
this article (see Ref. \cite{ShibataZPhysB}).

Since Eq.~(\ref{eq:LvN}) is in general a system of (infinitely) many
differential equations, exact solutions are only known in rare
cases. Moreover, even if an exact solution can be found, one is
usually not interested in the dynamics of the environment but wants to
calculate the time evolution of system observables. Therefore, we seek
an approximate equation of motion for the reduced density matrix
$\rho(t)=\mbox{Tr}_{\rm R}\{W(t)\}$ of the open system. In this
section we will describe three different approaches to that goal: the
Born-Markov approximation, the Nakajima-Zwanzig projection operator
technique and the time-convolutionless projection operator technique.

\subsection{The Markovian quantum master equation}
\label{sec:QME}

In this section we sketch an intuitive derivation of the
Markovian quantum master equation based on the Born-Markov
approximation (see, e.~g., \cite{Louisell,GardinerQN}).
The starting point is an exact equation of motion for the reduced
density matrix which can be obtained by integrating the
Liouville-von Neumann equation (\ref{eq:LvN}) twice, differentiating
with respect to $t$ and taking the trace over the reservoir. This
yields the exact equation of motion
\begin{equation}
  \label{eq:ex_mo}
  \dot\rho(t)=-\alpha^2\int_0^tds\,\mbox{Tr}_{\rm R}\left\{\left[H_{\rm I}(t),
  \left[ H_{\rm I}(s),W(s)\right]\right]\right\},
\end{equation}
which still contains the density matrix $W(t)$ of the composed
system. 

The first approximation we make is the Born approximation which
consists in approximating the density matrix of the composed system by
a product of the form
\begin{equation}
  \label{eq:Born}
  W(t)\approx\rho(t)\otimes\rho_{\rm R},
\end{equation}
where $\rho(t)$ refers to the variables of the reduced system and 
$\rho_{\rm R}$ denotes a stationary state of the environment. 
Such an approximation is justified if the coupling between the system and 
the environment is weak. Inserting Eq.~(\ref{eq:Born}) into
Eq.~(\ref{eq:ex_mo}) we obtain the closed integro-differential equation
for the reduced density matrix
\begin{equation}
  \label{eq:i_d_mo}
  \dot\rho(t)=-\alpha^2\int_0^tds\,\mbox{Tr}_{\rm R}\left\{\left[H_{\rm I}(t),
  \left[ H_{\rm I}(s),\rho(s)\otimes\rho_{\rm R}\right]\right]\right\}.
\end{equation}
This equation is further simplified by making the Markov
approximation: replacing $\rho(s)$ by $\rho(t)$ yields a closed differential
equation of motion for the reduced density matrix which contains only
$\rho(t)$, namely 
\begin{equation}
  \label{eq:mark_mo}
  \dot\rho(t)=-\alpha^2\int_0^tds\,\mbox{Tr}_{\rm R}\left\{\left[H_{\rm I}(t),
  \left[ H_{\rm I}(s),\rho(t)\otimes\rho_{\rm R}\right]\right]\right\}.
\end{equation}
The Markov approximation is based on the assumption that the
correlation time of the reservoir $\tau_{\rm R}$ is small compared to
the time scale $\tau_{\rm S}$ on which $\rho(t)$ changes. The final form of the
quantum master equation is obtained by extending the upper limit of
the integral to infinity, which is valid for times $t\gg\tau_{\rm R}$
since the integrand is negligible for $s\gg \tau_{\rm R}$.

Within this derivation of the quantum master equation, the Markov
approximation appears as an additional approximation besides the Born
approximation, and one is tempted to believe, that the generalized
master equation (\ref{eq:i_d_mo}) is more accurate than the master
equation (\ref{eq:mark_mo}). However, as we will see in
Secs.~\ref{sec:Naka} and \ref{sec:tcl}, both approximations are only
valid to second order in the coupling strength $\alpha$ and are hence
equally accurate (see also \cite{Kampenkumm1,vanKampen}). We will also
demonstrate this by means of a specific example in
Sec.~\ref{sec:dJCres}.

\subsection{Nakajima-Zwanzig projection operator technique}
\label{sec:Naka}

The Nakajima-Zwanzig projection operator technique
\cite{Nakajima,Zwanzig,Resibois} is based on a partition of the state
of a system into a relevant and an irrelevant part by defining an
adequate projection operator ${\cal P}$ which projects the state on
the relevant part and a projector ${\cal Q}= 1-{\cal P}$ which
projects on the irrelevant part. For our system reservoir model we
define the projector ${\cal P}$ in the usual way as
\begin{equation}
  \label{eq:projector}
  {\cal P}W(t)=\mbox{Tr}_{\rm R}\left\{W(t)\right\}\otimes \rho_{\rm
  R}\equiv\rho(t)\otimes\rho_{\rm R},
\end{equation}
where $\rho_{\rm R}$ is a stationary state of the reservoir. The equation of
motion for the two components ${\cal P}W(t)$ and ${\cal Q}W(t)$ can be
obtained directly form the Liouville-von Neumann equation (\ref{eq:LvN}):
\begin{eqnarray}
  \label{eq:P}
  \frac{\partial}{\partial t}{\cal P}W(t)&=&\alpha{\cal P}L(t){\cal P}W(t)+
  \alpha{\cal P}L(t){\cal Q}W(t)\\
  \label{eq:Q}
  \frac{\partial}{\partial t}{\cal Q}W(t)&=&\alpha{\cal Q}L(t){\cal P}W(t) 
  +\alpha{\cal Q}L(t){\cal Q}W(t).
\end{eqnarray}
Taking into account the initial condition Eq.~(\ref{eq:ini}) the formal solution of Eq.~(\ref{eq:Q}) reads
\begin{equation}
  \label{eq:Q_sol}
  {\cal Q}W(t)=\alpha\int_0^tds\,{\cal G}(t,s){\cal Q}L(s){\cal P}W(s)
\end{equation}
where ${\cal G}(t,s)$ is defined as
\begin{equation}
  \label{eq:G}
  {\cal G}(t,s)=T_\leftarrow\exp\left(\alpha\int_s^tds'\,{\cal Q}L(s')\right).
\end{equation}
The symbol $T_\leftarrow$ indicates the chronological time
ordering. Substituting the expression for ${\cal Q}W(t)$ into the
equation of motion of the relevant part of the state (\ref{eq:P}) we
obtain the generalized master equation for ${\cal P}W(t)$ 
\begin{equation}
  \label{eq:GME} 
  \frac{\partial}{\partial t}{\cal P}W(t)=\alpha {\cal P}L(t){\cal P}W(t)
  +\int_0^tds\,\widetilde K(t,s){\cal P}W(s),
\end{equation}
with the memory kernel
\begin{equation}
  \label{eq:Ktild}
  \widetilde K(t,s)=\alpha^2{\cal P}L(t){\cal G}(t,s){\cal Q}L(s).
\end{equation}
It is important to note, that the generalized master equation
(\ref{eq:GME}) is exact and that, hence, the explicit computation 
of the memory kernel $\widetilde K(t,s)$ is, in general, as
complicated as the explicit solution of the Liouville-von Neumann
equation (\ref{eq:LvN}). However, Eq.~(\ref{eq:GME}) 
serves as a starting point for systematic approximations.
For example, a perturbative expansion of the memory kernel $\widetilde
K(t,s)$ to second order in the coupling strength $\alpha$ leads to
the generalized quantum master equation in the Born approximation
(\ref{eq:i_d_mo}). On the other hand, although the computation of the
memory kernel is essentially facilitated by using a perturbative
expansion, the final form of the equation of motion is still an
integro-differential equation, the integration of which can be rather
difficult. We can overcome this by using the
time-convolutionless projection operator technique, which will
be described in the following Section.

\subsection{Time-convolutionless projection operator technique}
\label{sec:tcl}

The basic idea of the time-convolutionless projection operator
technique \cite{ShibataZPhysB,ShibataJStat} is to replace $W(s)$ in
the formal solution of the irrelevant part (\ref{eq:Q_sol}) by
\begin{equation}
  \label{eq:W}
  W(s)=G(t,s)({\cal P}+{\cal Q} )W(t),
\end{equation}
where the backward propagator $G(t,s)$ of the composite system is defined as
\begin{equation}
  \label{eq:G_back}
  G(t,s)=T_\rightarrow \exp\left(-\alpha\int_s^tds'\,L(s')\right),
\end{equation}
and $T_\rightarrow$ indicates the anti-chronological time
ordering. Solving Eq.~(\ref{eq:W}) for ${\cal Q}W(t)$, we find
\begin{equation}
  \label{eq:Q_new}
  {\cal Q}W(t)=\left[1-\Sigma(t)\right]^{-1}\Sigma(t){\cal P}W(t),
\end{equation}
with
\begin{equation}
  \label{eq:Sig}
  \Sigma(t)=\alpha\int_0^tds{\cal G}(t,s){\cal Q}L(s){\cal P}G(t,s),
\end{equation}
which can be substituted in Eq.~(\ref{eq:P}) to obtain the exact,
time-convolutionless equation of motion for the relevant part of the system
\begin{eqnarray}
  \label{eq:tcl_ex}
  \frac{\partial}{\partial t}{\cal P}W(t)&=&K(t){\cal P}W(t)\nonumber\\
  &\equiv&\alpha{\cal P}L(t)\left[1-\Sigma(t)\right]^{-1}{\cal P}W(t).
\end{eqnarray}
The crucial point of this construction is the existence of the
generator $K(t)$ which relies on the existence of the
operator $\left[1-\Sigma(t)\right]^{-1}$. Since $\Sigma(0)=0$ and
$\Sigma(t)$ is continuous, this operator exists for all $t$ if and
only if it can be expanded in a geometric series
\begin{equation}
  \label{eq:Sig_geo}
  \left[1-\Sigma(t)\right]^{-1}=\sum_{n=0}^\infty \Sigma(t)^n.
\end{equation}
This condition is always satisfied for short times or in the weak
coupling regime, but can be violated in the strong coupling
regime, as will be demonstrated explicitly in
Sec.~\ref{sec:dJCres}. Therefore, we define the intermediate coupling
regime as the range of coupling parameters $\alpha$, where
non-Markovian effects are significant, but the generator $K(t)$ exists
for all $t$.

Using Eq.~(\ref{eq:Sig_geo}) we can also write
the generator of the time-convolutionless master equation as 
\begin{equation}
  \label{eq:TCL_geo}
    K(t)=\sum_{n=0}^\infty\alpha{\cal P}L(t)\Sigma(t)^n.
\end{equation}
This form is the starting point for a perturbative expansion of $K(t)$
in powers of the coupling strength $\alpha$. To fourth order one
obtains, for example,
\begin{equation}
  \label{eq:K_pert}
  K(t)=\alpha^2K_2(t)+\alpha^4K_4(t)+{\cal O}(\alpha^6)
\end{equation}
where
\begin{equation}
  \label{eq:K_2}
  K_2(t)= \int_0^tdt_1{\cal P}L(t)L(t_1){\cal P},
\end{equation}
and 
\begin{eqnarray}
  \label{eq:K_4}
 \lefteqn{ K_4(t)  = \int_0^tdt_1\int_0^{t_1}dt_2\int_0^{t_2}
  dt_3}\\
  &&\times\Big[{\cal  P}L(t)L(t_1)L(t_2)L(t_3){\cal P}
  -{\cal  P}L(t)L(t_1){\cal  P}L(t_2)L(t_3){\cal P}\nonumber\\
  &&-{\cal  P}L(t)L(t_2){\cal  P}L(t_1)L(t_3){\cal P}
  -{\cal  P}L(t)L(t_3){\cal  P}L(t_1)L(t_2){\cal P}\Big] . \nonumber
\end{eqnarray}
The higher--order terms can be obtained in a way similar to van
Kampen's cumulant expansion \cite{Kampenkumm1,Kampenkumm2}.  All terms
containing odd orders of the coupling strength vanish in this
expansion, since by definition of ${\cal P}$ and $L(t)$ we have ${\cal
P}L(t_1)\cdots L(t_{2k+1}){\cal P}=0$ (see Eq.~(\ref{eq:odd})).  It is
important to note that the general structure of the
time-convolutionless equation of motion (\ref{eq:tcl_ex}) of the
reduced density matrix is not changed by the perturbative expansion,
i.~e., the approximative equation of motion is also linear in
$\rho(t)$ and local in time, unlike the perturbative expansion of the
generalized master equation (\ref{eq:GME}).

\subsubsection{The equation of motion to second order}
\label{sec:tcl_sec}

When we substitute the expressions for the generator $L(t)$ and the
projection operator ${\cal P}$, Eqs.~(\ref{eq:LvN}) and
(\ref{eq:projector}), respectively, into the second order contribution
to $K(t)$ we immediately obtain the time-dependent quantum master
equation (\ref{eq:mark_mo}) within the Born-Markov approximation
(without extending the upper limit of the time integration to infinity). Thus,
Eq.~(\ref{eq:i_d_mo}) as well as Eq.~(\ref{eq:mark_mo}) are correct to
the same order in the coupling and the major approximation in the
heuristic derivation of the quantum master equation in
Sec.~\ref{sec:QME} is not the Markov, but the Born approximation. This
seems to be somewhat counterintuitive, since after making the Born
approximation the equation of motion of the reduced density matrix is
still a complicated integro-differential equation, whereas the Markov
approximation considerably simplifies the calculations. Nevertheless,
it does in general not improve the accuracy of a calculation to make only the
Born-approximation and to omit the Markov-approximation.

\subsubsection{The equation of motion to fourth order}
\label{sec:tcl_for}

We now compute the explicit expression $K_4(t)$ for the fourth order
contribution to the time-convolutionless equation of motion. To this
end, we decompose the interaction Hamiltonian into a sum of products
in the form
\begin{equation}
  \label{eq:H_eig}
  H_{\rm I} = \sum_kF_k\otimes Q_k.
\end{equation}
We further assume, that the state $\rho_{\rm R}$ is not only
stationary, but also Gaussian, i.~e., 
\begin{eqnarray}
  \label{eq:Gauß}
  \lefteqn{\mbox{Tr}_{\rm R}\left\{\rho_{\rm R}Q_{i_0}(t)
  Q_{i_1}(t_1)Q_{i_2}(t_2)Q_{i_3}(t_3) \right\} =}\nonumber\\ 
 && \mbox{Tr}_{\rm R}\left\{\rho_{\rm R}Q_{i_0}(t) Q_{i_1}(t_1)\right\}
  \mbox{Tr}_{\rm R}\left\{\rho_{\rm R}Q_{i_2}(t_2) Q_{i_3}(t_3)\right\}\nonumber\\
 &+& \mbox{Tr}_{\rm R}\left\{\rho_{\rm R}Q_{i_0}(t) Q_{i_2}(t_2)\right\}
  \mbox{Tr}_{\rm R}\left\{\rho_{\rm R}Q_{i_1}(t_1) Q_{i_3}(t_3)\right\}\nonumber\\
 &+& \mbox{Tr}_{\rm R}\left\{\rho_{\rm R}Q_{i_0}(t) Q_{i_3}(t_3)\right\}
  \mbox{Tr}_{\rm R}\left\{\rho_{\rm R}Q_{i_1}(t_1) Q_{i_2}(t_2)\right\},
\end{eqnarray}
and we introduce the short-hand notation
\begin{eqnarray*}
  \hat 0, \hat 1, \cdots && \qquad\mbox{ denotes } F_{i_0}(t), F_{i_1}(t_1),
  \cdots\\
  \langle 1 2\rangle,\cdots&& \qquad\mbox{ denotes } \mbox{Tr}_{\rm R}\left\{
  \rho_{\rm R}Q_{i_1}(t_1)Q_{i_2}(t_2)\right\}, \cdots
\end{eqnarray*}
and sum over repeated indices $i_k$. In this notation we find for
example
\begin{eqnarray}
  \label{eq:not_examp}
  \lefteqn{{\cal P}L(t)\cdots L(t_3){\cal P}W(t)}\\
  &=&\left[\hat 0,\left[\cdots
  ,\left[\hat 3,\rho\right]\cdots\right]\right]\langle 0\cdots
  3\rangle\nonumber\\ 
  &=&\sum_{i_0\cdots i_3}\left[F_{i_0}(t),\left[\cdots,\left[
  F_{i_3}(t_3),\rho(t)\right]\cdots\right]\right]\nonumber\\
 &&\times \mbox{Tr}_{\rm R}\left\{\rho_{\rm R}Q_{i_0}(t)\cdots
  Q_{i_3}(t_3)\right\}\nonumber,
\end{eqnarray}
and inserting the expression (\ref{eq:H_eig}) into Eq.~(\ref{eq:K_4}),
we obtain
\begin{eqnarray}
  \label{eq:K_4_exp}
   \lefteqn{ K_4(t){\cal P}W(t)=\int_0^tdt_1\int_0^{t_1}dt_2\int_0^{t_2}
  dt_3}\\
  &&\times\big\{\langle 02\rangle\langle 13\rangle\left[\hat 0,\left[\hat 1,\hat 2\right]\hat 3\rho\right] 
   - \langle 02\rangle\langle 31\rangle\left[\hat 0,\left[\hat 1,\hat 2\right]\rho \hat 3\right]\nonumber\\
 &&  - \langle 20\rangle\langle 13\rangle\left[\hat 0,\hat 3\rho\left[\hat 1,\hat 2\right]\right]
   + \langle 20\rangle\langle 31\rangle\left[\hat 0,\rho \hat 3\left[\hat 1,\hat 2\right]\right]\nonumber\\
 &&  + \langle 03\rangle\langle
  12\rangle\left(\left[\hat 0,\left[\hat 3,\hat 2\right]\rho \hat 1\right]+\left[\hat 0,\left[\hat 1\hat 2,\hat 3\right]\rho\right]\right)\nonumber\\
 &&  + \langle 30\rangle\langle 21\rangle\left(\left[\hat 0,\hat 1\rho\left[\hat 3,\hat 2\right]\right]+\left[\hat 0,\rho\left[\hat 2\hat 1,\hat 3\right]\right]\right)\nonumber\\
 &&  - \langle 03\rangle\langle 21\rangle\left[\hat 0,\left[\hat 1,\hat 3\right]\rho\hat 2\right]
   - \langle 30 \rangle\langle12\rangle\left[\hat 0,\hat 2\rho\left[\hat 1,\hat 3\right]\right]\big\}\nonumber.
\end{eqnarray}
Note that this expression contains commutators between various system
operators, which can immensely simplify the explicit evaluation of
$K_4(t)$, if certain commutation relations are specified, such as
bosonic commutation relations for a harmonic oscillator, or the
commutation relations for the pseudospin operators (see Sec.~\ref{sec:eq_mo}).

\section{Stochastic unravelling of quantum master equations}
\label{sec:stqme}

\subsection{Quantum master equations in Lindblad form}
\label{sec:stlind}

In Ref. \cite{Lindblad} Lindblad has shown that the equation of
motion of a reduced density matrix has to be of the form 
\begin{eqnarray}
  \label{eq:QME}
  \lefteqn{\frac{\partial}{\partial
  t}\rho(t)=-i\left[H_{\rm S}+\frac{1}{2}\sum_iS_i(t)L_i^\dagger L_i,\rho(t)\right]}\\
  &+&\sum_i\gamma_i(t)\left\{-\frac{1}{2}L_i^\dagger L_i\rho(t)
  -\frac{1}{2}\rho(t)L_i^\dagger L_i+L_i\rho(t)L_i^\dagger\right\}, \nonumber
\end{eqnarray}
if the dynamics of the reduced system is assumed to conserve positivity and to 
represent a quantum dynamical semi-group. Here, $H_{\rm S}$ is the
Hamiltonian of the system, the time-dependent coefficients $S_i(t)$
describe an energy shift induced by the coupling to the environment,
namely the Lamb and Stark shifts, and the positive rates $\gamma_i(t)$
model the dissipative coupling to the $i-$th decay channel.

In this case, the state of the open system can alternatively be
described by a stochastic wave function $\psi(t)$
\cite{Carmichael,MolmerPRL68,GardinerPRA46,ZollerPRA46,Gisin:92,Gisin:93},
the covariance matrix of which equals the reduced density matrix,
i.~e.,
\begin{equation}
  \label{eq:P_rho}
  \rho(t)=\int D\psi D\psi^*\,|\psi\rangle\langle\psi| P[\psi,t],
\end{equation}
where $P[\psi,t]$ is the probability density functional of finding the
state of the open system in the Hilbert space volume element $D\psi
D\psi^*$ at the time $t$ \cite{BP:QS3,BP:QS4}.

The time evolution of the stochastic wave function is governed by a
stochastic differential equation, which might either be diffusive
\cite{Gisin:92,Gisin:93} or of the piecewise deterministic jump type
\cite{Carmichael,MolmerPRL68,GardinerPRA46,ZollerPRA46}. The latter
takes the form \cite{WisemanPRA93}
\begin{equation}
  \label{eq:SDE}
  d\psi(t)=-iG(\psi,t) dt+
  \sum_i\left(\frac{L_i\psi(t)}{\|L_i\psi(t)\|}-\psi(t)\right)dN_i(t),
\end{equation}
where the $dN_i(t)$ are the differentials of independent Poisson
process $N_i(t)$ with mean $\langle
dN_i(t)\rangle=\gamma_i(t)\|L_i\psi(t)\|^2dt$. The drift generator
takes the form
\begin{eqnarray}
  \label{eq:G_eq}
  G(\psi,t)&=&H(t)\psi+\frac{1}{2}\sum_iS_i(t) L_i^\dagger L_i\psi\nonumber\\
  &&-\frac{i}{2}\sum_i\gamma_i(t)\left(L_i^\dagger L_i -\|L_i\psi\|^2 
  \right)\psi.
\end{eqnarray}
For the differential of the Poisson process $dN_i(t)$ the Ito rule
$dN_i(t)dN_j(t)=\delta_{ij}dN_i(t)$ holds, that is, $dN_i(t)$ can either
be $0$ or $1$. If $dN_i(t)=0$, then the system evolves continuously
according to the nonlinear Schr\"odinger-type equation
\begin{equation}
  \label{eq:cont}
  i\frac{\partial}{\partial t}\psi(t)=G(\psi,t),
\end{equation}
whereas, if $dN_i(t)=1$ for some $i$, then the system undergoes an
instantaneous, discontinuous transition of the form
\begin{equation}
  \label{eq:trans}
  \psi(t)\longrightarrow\frac{L_i\psi(t)}{\|L_i\psi(t)\|}.
\end{equation}
Note that the generator $G(\psi,t)$ of the continuous time evolution 
is non-Hermitian and hence the propagator of $\psi(t)$ is non-unitary.
However, due to the nonlinearity of the generator, the norm of
$\psi(t)$ is preserved in time.

Using the Ito calculus for the differentials $dN_i(t)$  it is
easy to check that the equation of motion of the covariance matrix of
$\psi(t)$ equals the usual Markovian quantum master equation
(\ref{eq:QME}) in Lindblad form. Thus, expectation values of system
observables can either be calculated by means of the reduced density
matrix or as averages over different realizations of the
stochastic process $\psi(t)$ and both descriptions yield the same
results. 

\subsection{General quantum master equations}
\label{sec:SGQME}

The most general type of a quantum master equation which results from
the time-convolutionless projection operator technique -- or from a
perturbative approximation -- is linear in $\rho(t)$ and local in time
(see Sec.~\ref{sec:tcl}) but needs not to be in the Lindblad form, as
we will show in an example below (see Sec.~\ref{sec:dJCdet}). However,
these equations can always be written in the form
\begin{equation}
  \label{eq:rho_mo}
  \frac{\partial}{\partial t}\rho(t)=A(t)\rho(t)+\rho(t)B^\dagger(t)+\sum_i
  C_i(t)\rho(t)D_i^\dagger(t),
\end{equation}
with some time-dependent linear operators $A(t)$, $B(t)$, $C_i(t)$,
and $D_i(t)$. In order to find an unraveling of this equation of
motion we follow a strategy, which has already been successfully
applied to the calculation of multitime correlation functions
\cite{BP:QS12,BP:QS14}. We describe the state of the open system by a
pair of stochastic wave functions 
\begin{equation}
\theta(t)= \left( \begin{array}{c}
                  \phi(t) \\
                  \psi(t)
                  \end{array} \right) .
\end{equation}
Formally, $\theta(t)$ can be regarded
as an element of the doubled Hilbert space $\widetilde{\cal H}={\cal
H}\oplus{\cal H}$. If $\widetilde P[\theta,t]$ denotes the probability
density functional of the process in the doubled Hilbert space
$\widetilde{\cal H}$, we may define the reduced density matrix as 
\begin{equation}
  \label{eq:rho_dbl_P}
  \rho(t)=\int D\theta D\theta^* |\phi\rangle\langle\psi|\widetilde
  P[\theta,t].
\end{equation}
The time evolution of the state vector
$\theta(t)$ is then governed by the stochastic differential equation
\begin{eqnarray}
  \label{eq:gen_un}
  d\theta(t)&=&-iG(\theta,t)dt\\
  &&+\sum_i\left(\frac{\|\theta(t)\|}{\left\|
  J_i(t)\theta(t)\right\|} 
  J_i(t)\theta(t)-\theta(t)\right)dN_i(t) \nonumber,
\end{eqnarray}
where $dN_i(t)$ is the differential of a Poisson process with mean
\begin{equation}
  \label{eq:dbl_mean}
  \langle dN_i(t)\rangle=\frac{\left\| J_i(t)\theta(t)\right\|^2}
  {\|\theta(t)\|^2}dt,
\end{equation}
and the functional $G(\theta,t)$ is defined as 
\begin{equation}
  \label{eq:G_dbl}
   G(\theta,t)=i\left( F(t)+\frac{1}{2}\sum_i 
  \frac{\left\| J_i(t)\theta(t)\right\|^2}{\|\theta(t)\|^2}\right)
  \theta(t),
\end{equation}
with the time-dependent operators
\begin{equation}
  \label{eq:AB}
   F(t)=
  \left(\begin{array}{cc}
  A(t)&0\\
  0&B(t)\end{array}\right),\quad
   J_i(t)=
  \left(\begin{array}{cc}
  C_i(t)&0\\
  0&D_i(t)\end{array}\right).
\end{equation}
Again, this type of stochastic evolution equation describes a
piecewise deterministic jump process, where the deterministic pieces are
solutions of the differential equation
\begin{equation}
  \label{eq:cont_dbl}
  i\frac{\partial}{\partial t}\theta(t)=G(\theta,t),
\end{equation}
and the jumps induce transitions of the form
\begin{equation}
  \label{eq:trans_dbl}
  \theta(t)\longrightarrow\frac{\|\theta(t)\|}{\|J_i\theta(t)\|}J_i\theta(t)
  =\frac{\|\theta(t)\|}{\|J_i\theta(t)\|}\left(
    \begin{array}{c}
      C_i\phi\\D_i\psi
    \end{array}
\right).
\end{equation}
Note, that the structure of the stochastic differential equation in the
doubled Hilbert space (\ref{eq:gen_un}) is very similar to
the structure of the stochastic differential equation
(\ref{eq:SDE}). In fact, the unraveling of general quantum
master equations presented in this section contains as a special case the
unraveling of Lindblad--type equations shown in Sec.~\ref{sec:stlind}: If we
set 
\begin{equation}
  \label{eq:A_B_eq}
  A(t)=B(t)=
  -iH_{\rm S}
  -\frac{1}{2}\sum_k\left[\gamma_k(t) +iS_k(t)\right] L_k^\dagger L_k
\end{equation}
and
\begin{equation}
  \label{eq:C_D}
  C_i(t)=D_i(t)=\sqrt{\gamma(t)}L_i,
\end{equation}
the equation of motion (\ref{eq:rho_mo}) reduces to the Lindblad
equation (\ref{eq:QME}) and both unravelings are identical. 

\section{Example: The spontaneous decay of a two-level system}
\label{sec:two-lev}

In this section we consider as an example of the general theory the
exactly solvable model of a two-level system spontaneously decaying
into the vacuum within the rotating wave approximation. The
Hamiltonian of the total system is given by
\begin{eqnarray}
  \label{eq:H_0}
  H_{\rm 0}&=&\omega_{\rm S}\sigma^+\sigma^-+\sum_k\omega_k b_k^\dagger b_k,\\
  \label{eq:H_I}
  H_{\rm I}&=&\sigma^+\otimes B+\sigma^-\otimes B^\dagger \mbox{ with
  } B=\sum_k g_k b_k,
\end{eqnarray}
where $\omega_{\rm S}$ denotes the transition frequency of the
two-level system, the index $k$ labels the different field modes with
frequency $\omega_k$, annihilation operator $b_k$ and coupling constant
$g_k$, and $\sigma^\pm$ denote the pseudospin operators. 

\subsection{Exact and approximated equations of motions}
\label{sec:eq_mo}

The exact solution and equation of motion for this model can be
obtained in the following way: Define the states \cite{GarrawayPRA55}
\begin{eqnarray}
  \label{eq:expand}
  \psi_0&=&|0\rangle_{\rm S}\otimes|0\rangle_{\rm R}\nonumber\\
  \psi_1&=&|1\rangle_{\rm S}\otimes|0\rangle_{\rm R}\nonumber\\
  \psi_k&=&|0\rangle_{\rm S}\otimes|k\rangle_{\rm R}
\end{eqnarray}
where $|0\rangle_{\rm S}$ and $|1\rangle_{\rm S}$ indicate the ground
and excited state of the system, respectively, the state
$|0\rangle_{\rm R}$ denotes the vacuum state of the reservoir, and
$|k\rangle_{\rm R}=b_k^\dagger|0\rangle_{\rm R}$ denotes the state
with one photon in mode $k$. Since the interaction Hamiltonian
conserves the total number of particles, the flow of the Schr\"odinger
equation generated by $H_{\rm I}$ is confined to the
subspace spanned by these vectors. Hence, we may expand the state of
the total system at any time as 
\begin{equation}
  \label{eq:phi_exp}
  \phi(t)=c_0\psi_0+c_1(t)\psi_1+\sum_k c_k(t)\psi_k,
\end{equation}
with some probability amplitudes $c_0$, $c_1(t)$, and $c_k(t)$. The
time evolution of these probability amplitudes is determined by a
complicated system of ordinary differential equations, which can be
solved in some simple cases by introducing the so-called pseudomodes
\cite{GarrawayPRA55}. With these probability amplitudes, the reduced
density matrix takes the form
\begin{equation}
  \label{eq:rho_ex}
  \rho(t)=\left(\begin{array}{cc}
  |c_1(t)|^2& c_1(t) c_0^*\\
  c_1^*(t)c_0& |c_0|^2+\sum_k|c_k(t)|^2
  \end{array}\right) .
\end{equation}
Differentiating this expression with respect to time we get
the following exact equation of motion,
\begin{eqnarray}
  \label{eq:TCL-QME}
  \lefteqn{\frac{\partial}{\partial t}\rho(t)=-\frac{i}{2}S(t)
  [\sigma^+\sigma^-,\rho(t)]}\\
  &&+\gamma(t)\left\{-\frac{1}{2}\sigma^+\sigma^-\rho(t)
  -\frac{1}{2}\rho(t)\sigma^+\sigma^- +\sigma^-\rho(t)\sigma^+
  \right\},\nonumber
\end{eqnarray}
where the time-dependent energy shift $S(t)$ and decay rate $\gamma(t)$ are
defined as 
\begin{equation}
  \label{eq:S_gamm}
  S(t)=-2\Im\left\{\frac{\dot c_1(t)}{c_1(t)}\right\},\quad
  \gamma(t)=-2\Re\left\{\frac{\dot c_1(t)}{c_1(t)}\right\}.
\end{equation}
Note, that if the decay rate $\gamma(t)$ is positive for all $t$, then
this equation of motion is in the Lindblad form (\ref{eq:QME}).

The equation of motion within the Born approximation can be expressed
in terms of the reservoir correlation function. To this end, we define
the real functions $\Phi(t)$ and $\Psi(t)$ as
\begin{eqnarray}
  \label{eq:phi_psi}
   \Phi(t)+i\Psi(t)&=&2 \mbox{Tr}_{\rm R}\left\{B(t)B^\dagger
   \rho_{\rm R}\right\}e^{i\omega_{\rm S}t}\nonumber\\ 
   &=&2\int d\omega J(\omega)e^{i(\omega_{\rm S}-\omega)t},
\end{eqnarray}
where $B(t)=\exp(iH_0t)B\exp(-iH_0t)$, and we have performed the
continuum limit. $J(\omega)$ is the spectral density.
The equation of motion in the Born
approximation~(\ref{eq:i_d_mo}) then reads
\begin{eqnarray}
  \label{eq:Borntl}
  \lefteqn{\frac{\partial}{\partial t}\rho(t)=-\int_0^tds\,\bigg\{
  \frac{i}{2}\Psi(t-s)[\sigma^+\sigma^-,\rho(s)]}\\
  &&+\Phi(t-s)\left[\frac{1}{2}\sigma^+\sigma^-\rho(s)
  +\frac{1}{2}\rho(s)\sigma^+\sigma^- -\sigma^-\rho(s)\sigma^+
  \right]\bigg\}.\nonumber
\end{eqnarray}
Performing the Markov approximation and extending the upper limit of
the time integral to infinity, we obtain the usual time-independent
quantum master equation
\begin{eqnarray}
  \label{eq:MArktl}
  \lefteqn{\frac{\partial}{\partial t}\rho(t)=-\frac{i}{2}S_{\rm M}
  [\sigma^+\sigma^-,\rho(t)]}\\
  &&+\gamma_{\rm M}\left\{-\frac{1}{2}\sigma^+\sigma^-\rho(t)
  -\frac{1}{2}\rho(t)\sigma^+\sigma^- +\sigma^-\rho(t)\sigma^+
  \right\},\nonumber
\end{eqnarray}
where the Markovian Lamb shift $S_{\rm M}$ and the Markovian decay rate 
$\gamma_{\rm M}$ are defined as
\begin{equation}
  \label{eq:Mark_gamma_S}
  S_{\rm M}=\int_0^\infty ds\,\Psi(s), \;\;\;
  \gamma_{\rm M}=\int_0^\infty ds\,\Phi(s).
\end{equation}

The time-convolutionless expansion of the equation of motion according
to Sec.~\ref{sec:tcl} leads to a quantum master equation which has the
same structure as the exact equation of motion, but the time-dependent
energy shift $S(t)$ and decay rate $\gamma(t)$ are approximated by the
quantities
\begin{eqnarray}
  \label{eq:Lamb}
 \lefteqn{ S^{(4)}(t)=\int_0^tdt_1\Psi(t-t_1)+\frac{1}{2}
  \int_0^tdt_1\int_0^{t_1}dt_2\int_0^{t_2}dt_3 }\nonumber\\
  &\times&\Big[\Psi(t-t_2)\Phi(t_1-t_3)+\Phi(t-t_2)\Psi(t_1-t_3)\nonumber\\
  &+&\Psi(t-t_3)\Phi(t_1-t_2)+\Phi(t-t_3)\Psi(t_1-t_2)\Big]
\end{eqnarray}
and
\begin{eqnarray}
  \label{eq:rate}
  \lefteqn{\gamma^{(4)}(t)=\int_0^tdt_1\Phi(t-t_1)+\frac{1}{2}\int_0^tdt_1\int_0^{t_1}
  dt_2\int_0^{t_2} dt_3}\nonumber\\ 
  &\times&\Big[\Psi(t-t_2)\Psi(t_1-t_3)-\Phi(t-t_2)\Phi(t_1-t_3)\nonumber\\
  &+&\Psi(t-t_3)\Psi(t_1-t_2)-\Phi(t-t_3)\Phi(t_1-t_2)\Big].
\end{eqnarray}
It is important to note that the explicit expressions for $S^{(4)}(t)$
and $\gamma^{(4)}(t)$ only involve ordinary integrations over the
reservoir correlation functions, which can be done analytically
in simple cases or numerically.

\subsection{The damped Jaynes-Cummings model on resonance}
\label{sec:dJCres}

\begin{figure}[t]
  \begin{center}
    \leavevmode \epsfxsize.49\linewidth\epsffile{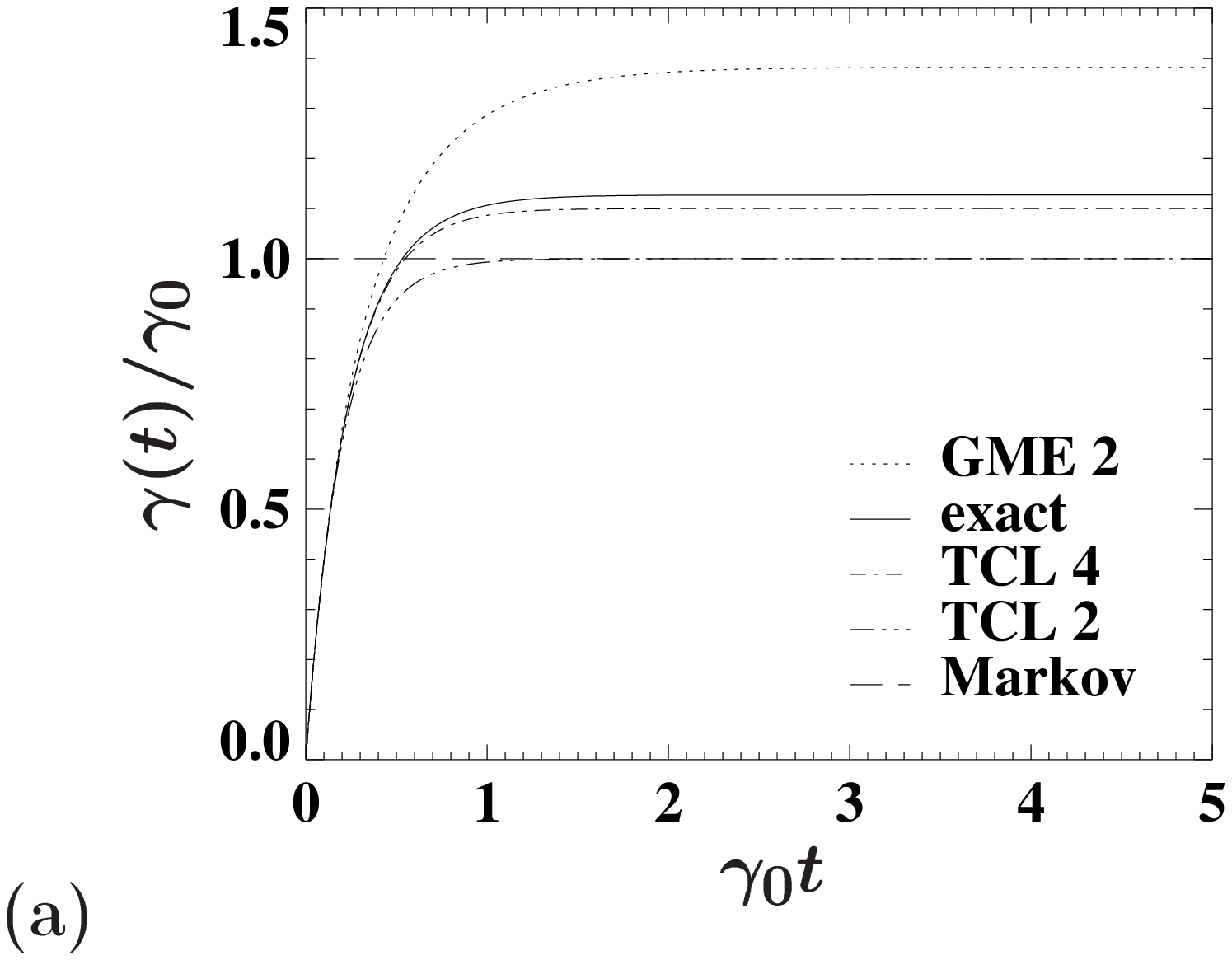}
    \leavevmode \epsfxsize.49\linewidth\epsffile{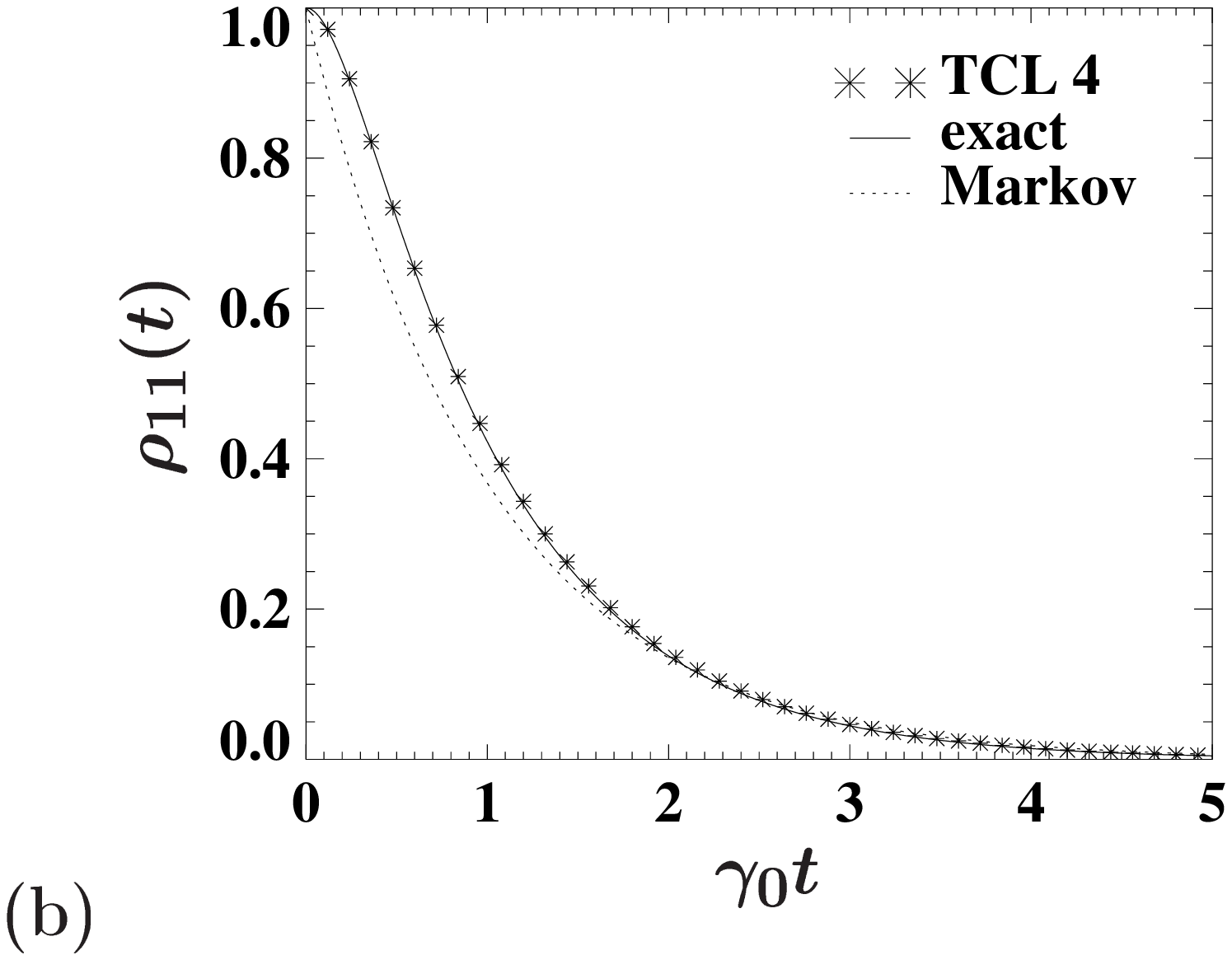}\\
    \leavevmode \epsfxsize.49\linewidth\epsffile{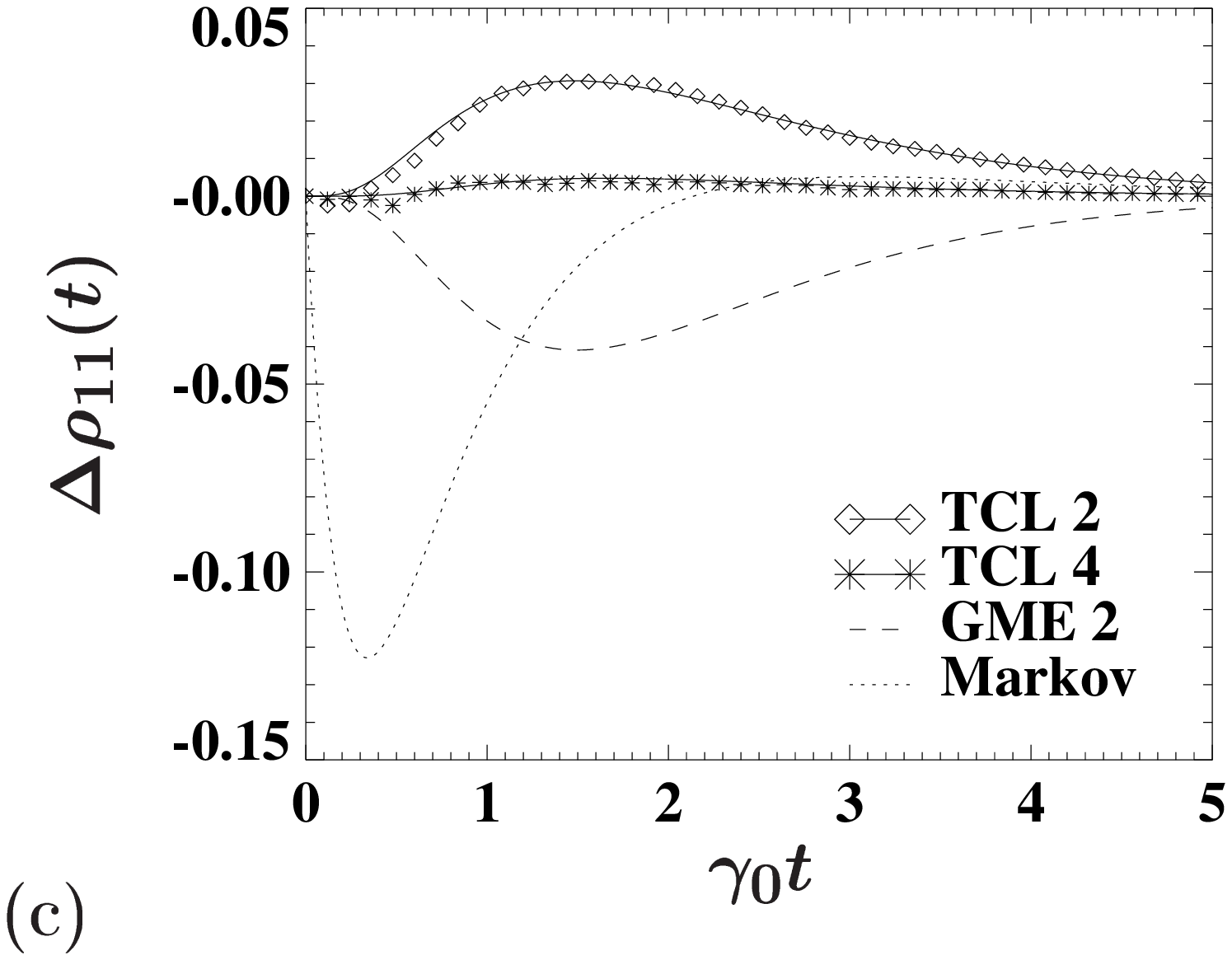} 
    \leavevmode \epsfxsize.49\linewidth\epsffile{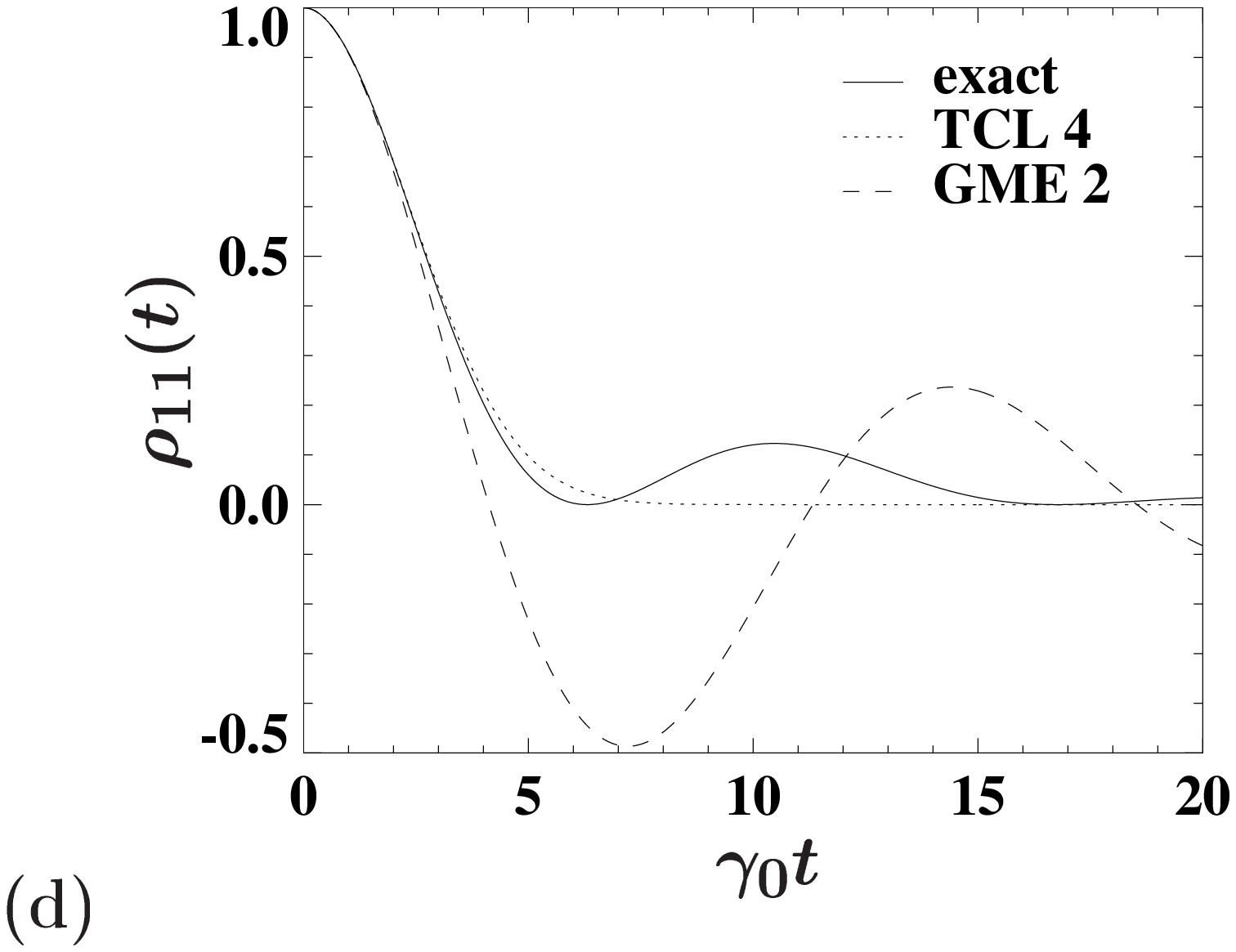}
    \caption{The damped Jaynes-Cummings model on resonance. Exact
      solution (exact), time-convolutionless master equation to second
      (TCL 2) and fourth order (TCL 4), generalized master equation to
      second order (GME 2), and the Markovian quantum master equation
      (Markov): (a) Decay rate of the excited state population, (b)
      the population of the excited state, including a stochastic
      simulation of the time-convolutionless quantum master equation
      with $10^5$ realizations, and (c) deviation of the approximative
      solutions from the exact result, for $\tau_{\rm S}=5\tau_{\rm
      R}$ (moderate coupling). (d) Population of the excited state for
      $\tau_{\rm S}=0.2\tau_{\rm R}$ (strong coupling).}
  \label{fig:1}
  \end{center}
\end{figure}

The damped Jaynes Cummings model describes the coupling of a two-level
atom to a single cavity mode which in turn is coupled to a reservoir
consisting of harmonic oscillators in the vacuum state. If we restrict
ourselves to the case of a single excitation in the atom--cavity
system, the cavity mode can be eliminated in favor of an effective
spectral density of the form
\begin{equation}
  \label{eq:JC}
  J(\omega)=\frac{1}{2\pi}\frac{\gamma_0\lambda^2} 
  {(\omega_{\rm S}-\omega)^2+\lambda^2},
\end{equation}
where $\omega_{\rm S}$ is the transition frequency of the two-level
system. The parameter $\lambda$ defines the spectral width of the
coupling, which is connected to the reservoir correlation time
$\tau_{\rm R}$ by the relation $\tau_{\rm R}=\lambda^{-1}$ and the
time scale $\tau_{\rm S}$ on which the state of the system changes is
given by $\tau_{\rm S}=\gamma_0^{-1}$. The exact probability amplitude
$c_1(t)$ (see Eq.~\ref{eq:phi_exp})) is readily obtained by using the
method of poles \cite{GarrawayPRA55}, since $J(\omega)$ has simple
poles at $\omega=\omega_0\pm i\lambda$. One gets
\begin{equation}
  \label{eq:c_1}
  c_1(t)=c_1(0)e^{-\lambda t/2}\left(\cosh\frac{dt}{2}+
  \frac{\lambda}{d}\sinh\frac{dt}2{}\right),
\end{equation}
where $d=\sqrt{\lambda^2-2\gamma_0\lambda}$, which yields the
time-dependent population of the excited state
\begin{equation}
  \label{eq:rho_11_ex}
  \rho_{11}(t)=\rho_{11}(0)e^{-\lambda t}\left(\cosh\frac{dt}{2}+
  \frac{\lambda}{d}\sinh\frac{dt}{2}\right)^2.
\end{equation}
Using Eq.~(\ref{eq:S_gamm}) we therefore obtain a vanishing
Lamb shift, $S(t)\equiv 0$, and the time-dependent decay rate
\begin{equation}
  \label{eq:gamma_ex}
  \gamma(t)=\frac{2\gamma_0\lambda\sinh({dt}/{2})}
  {d\cosh({dt}/{2})+\lambda\sinh({dt}/{2})}.
\end{equation}
In Fig.~\ref{fig:1} (a) we illustrate this time-dependent decay rate
$\gamma(t)$ ('exact') together with the Markovian decay rate
$\gamma_{\rm M}=\gamma_0$ ('Markov') for $\tau_{\rm S}=5\tau_{\rm R}$.
Note that for short times, i.~e., for times of the order of $\tau_{\rm
R}$, the exact decay rate grows linearly in $t$, which leads to the
quantum-mechanically correct short--time behavior of the transition
probability. In the long-time limit the decay rate saturates
at a value larger than the Markovian decay rate, which represents
corrections to the golden rule. The population of the excited state is
depicted in Fig.~\ref{fig:1} (b): for short times, the exact
population decreases quadratically and is larger than the Markovian
population, which is simply given by $\rho_{11}(0)\exp\left(-\gamma_0
t\right)$, whereas in the long-time limit the exact population is
slightly less than the Markovian population.

Next, we want to determine the solution of the generalized quantum
master equation in the Born approximation. To this end, we insert the
spectral density of the coupling strength (\ref{eq:JC}) into
Eq.~(\ref{eq:phi_psi}) to obtain $\Psi(t)\equiv 0$ and
\begin{equation}
  \label{eq:exp_kern}
  \Phi(t)=\gamma_0\lambda\exp(-\lambda t).
\end{equation}
The solution of the generalized master equation (\ref{eq:Borntl}) can
be found in the following way. We differentiate Eq.~(\ref{eq:Borntl})
with respect to $t$ and obtain
\begin{eqnarray}
  \label{eq:Born_exp}
  \ddot{\rho}(t)&=&-\lambda\dot\rho(t)\\
&+&\gamma_0\lambda\left[-\frac{1}{2}\sigma^+\sigma^-\rho(t)
  -\frac{1}{2}\rho(t)\sigma^+\sigma^- +\sigma^-\rho(t)\sigma^+
  \right].\nonumber
\end{eqnarray}
Due to the exponential memory kernel, this equation of motion is an
ordinary differential equation which is local in time, and contains
only $\rho(t)$, $\dot\rho(t)$ and $\ddot\rho(t)$. Solving this system
of differential equations for $\rho(t)$, we obtain the time evolution
of the population of the upper level 
\begin{equation}
  \label{eq:rho_GME}
  \widetilde\rho_{11}(t)=\rho_{11}(0)e^{-\lambda t/2}\left(\cosh\frac{d't}{2}+
  \frac{\lambda}{d'}\sinh\frac{d't}{2}\right),
\end{equation}
where $d'=\sqrt{\lambda^2-4\gamma_0\lambda}$. From this expression, we
can determine the time-dependent decay rate
\begin{equation}
  \label{eq:GME_rate}
  \widetilde\gamma(t)=\frac{\dot\rho_{11}(t)}{\rho_{11}(t)}= 
  \frac{2\gamma_0\lambda\sinh({d't}/{2})}
  {d'\cosh({d't}/{2})+\lambda\sinh({d't}/{2})},
\end{equation}
the structure of which is similar to the exact decay rate
(\ref{eq:gamma_ex}). Note, however, the difference between the
parameters $d$ and $d'$ which can also be seen in Fig~\ref{fig:1} (a)
where we have also plotted the decay rate $\widetilde\gamma(t)$ ('GME
2'): For short times, the decay rate $\widetilde\gamma(t)$ is in good
agreement with $\gamma(t)$, but in the long time limit,
$\widetilde\gamma(t)$ is too large.

Finally, the time-convolutionless decay rate can be determined from
Eq.~(\ref{eq:rate}), and to second and fourth order in the coupling we
obtain
\begin{equation}
  \label{eq:gamm2TCL}
  \gamma^{(2)}(t)=\gamma_0\left(1-e^{-\lambda t}\right),
\end{equation}
and
\begin{equation}
  \label{eq:gamm4TCL}
  \gamma^{(4)}(t)=\gamma_0\left\{1-e^{-\lambda t}+\frac{\gamma_0}{\lambda}
  \left[\sinh(\lambda t)-\lambda t\right]e^{-\lambda t}\right\},
\end{equation}
respectively, which corresponds to a Taylor expansion of the exact
decay rate $\gamma(t)$ in powers of $\gamma_0$, as can be checked by
differentiating $\gamma(t)$ with respect to $\gamma_0$.
Fig.~\ref{fig:1} (a) clearly shows, that $\gamma^{(2)}(t)$ as well as
$\gamma^{(4)}(t)$ approximate the exact decay rate very good for short
times, and $\gamma^{(4)}(t)$ is also a good approximation in the long
time limit. 

The time evolution of the population of the excited state can be
obtained by integrating the rate $\gamma^{(4)}(t)$ with respect to
$t$. This yields
\begin{equation}
  \label{eq:rho_4}
  \rho^{(4)}_{11}(t)=\rho_{11}(0)\exp\left(-\int_0^tds\,\gamma^{(4)}(s)\right).
\end{equation}
In order to compare the quality of the different approximation
schemes, we show the difference between the approximated populations
and the exact population in Fig.~\ref{fig:1} (c). Besides the
analytical solutions of the generalized master equation
(\ref{eq:Borntl}) and the time-convolutionless master equations, we
have also performed a stochastic simulation of the
time-convolutionless quantum master equations with $10^5$
realizations. Since the approximated rates $\gamma^{(2,4)}(t)$ are
positive for all $t$, the corresponding master equations are in
Lindblad form, and we can use the stochastic simulation algorithm
described in Sec.~\ref{sec:stlind} as an unravelling. Fig.~\ref{fig:1}
(c) shows, that the stochastic simulation is in very good agreement
with the corresponding analytical solutions. Moreover, we see that the
difference between the solution of the time-convolutionless master
equation to fourth order and the exact master equation is
small (see also Fig~\ref{fig:1} (b)), whereas the errors of the
generalized and the time-convolutionless master equation to second
order which correspond to the Born approximation and the Born-Markov
approximation (without extending the integral), respectively, are
larger and of the same order of magnitude. In fact, the Markov
approximation even leads to a slight improvement of the accuracy,
compared to the Born approximation, which is surprising if we consider
the heuristic derivation of the quantum master equation in
Sec.~\ref{sec:QME}.

As we pointed out in Sec.~\ref{sec:GME}, the approximation schemes
used in this article are perturbative and hence rely on the assumption
that the coupling is not too strong. But what happens, if the system
approaches the strong coupling regime? We will investigate this
question by means of the damped Jaynes-Cummings model on resonance,
where the explicit expressions of the quantities of interest are
known.

First, let us take a look at the exact expression for the population
of the excited state (\ref{eq:rho_11_ex}): In the strong coupling
regime, i.~e., for $\gamma_0>\lambda/2$ or $\tau_{\rm S}<2\tau_{\rm
R}$, the parameter $d$ is purely imaginary. Defining $\hat d=-id$ we
can write the exact population as
\begin{equation}
  \label{eq:rho_11_ex_im}
  \rho_{11}(t)=\rho_{11}(0)e^{-\lambda t}\left(\cos\frac{\hat dt}{2}+
  \frac{\lambda}{d}\sin\frac{\hat dt}{2}\right)^2,
\end{equation}
which is an oscillating function that has discrete zeros at
\begin{equation}
  \label{eq:zeros}
  t=\frac{2}{\hat d}\left(\pi - \arctan\frac{\hat d}{\lambda}\right).
\end{equation}
Hence, the rate $\gamma(t)$ diverges at these points (see
Eq.~(\ref{eq:S_gamm})). Obviously, $\gamma(t)$ can only be an
analytical function for $t\in [0,t_0[$, where $t_0$ is the smallest
positive zero of $\rho_{11}(t)$.

On the other hand, as we have seen in Sec.~\ref{sec:dJCres}, the
time-convolutionless quantum master equation corresponds basically to
a Taylor expansion of $\gamma(t)$ in powers of $\gamma_0$, and the
radius of convergence of this series is given by the region of
analyticity of $\gamma(t)$. For $\gamma_0<\lambda/2$, this is the
whole positive real axis, but for $\gamma_0>\lambda/2$ the
perturbative expansion only converges for $t<t_0$. This behavior can
be clearly seen in Fig.~\ref{fig:1} (d), where we have depicted
$\rho_{11}(t)$ and $\rho_{11}^{(4)}(t)$ for $\tau_{\rm S}=\tau_{\rm
R}/5$, i.~e., for a strong coupling: the perturbative expansion
converges to $\rho_{11}(t)$ for $t\lesssim t_0\approx 6.3/\gamma_0$,
but fails to converge for $t>t_0$.

The solution of the generalized master equation to second order shows
a quite distinct behavior, but also fails in the strong coupling
regime: for $\gamma_0>\lambda/4$ the population $\rho_{11}(t)$ starts
to oscillate and even takes negative values, which is unphysical (see
Fig.~\ref{fig:1} (d)).

\begin{figure}[t]
  \begin{center}
    \leavevmode \epsfxsize.49\linewidth\epsffile{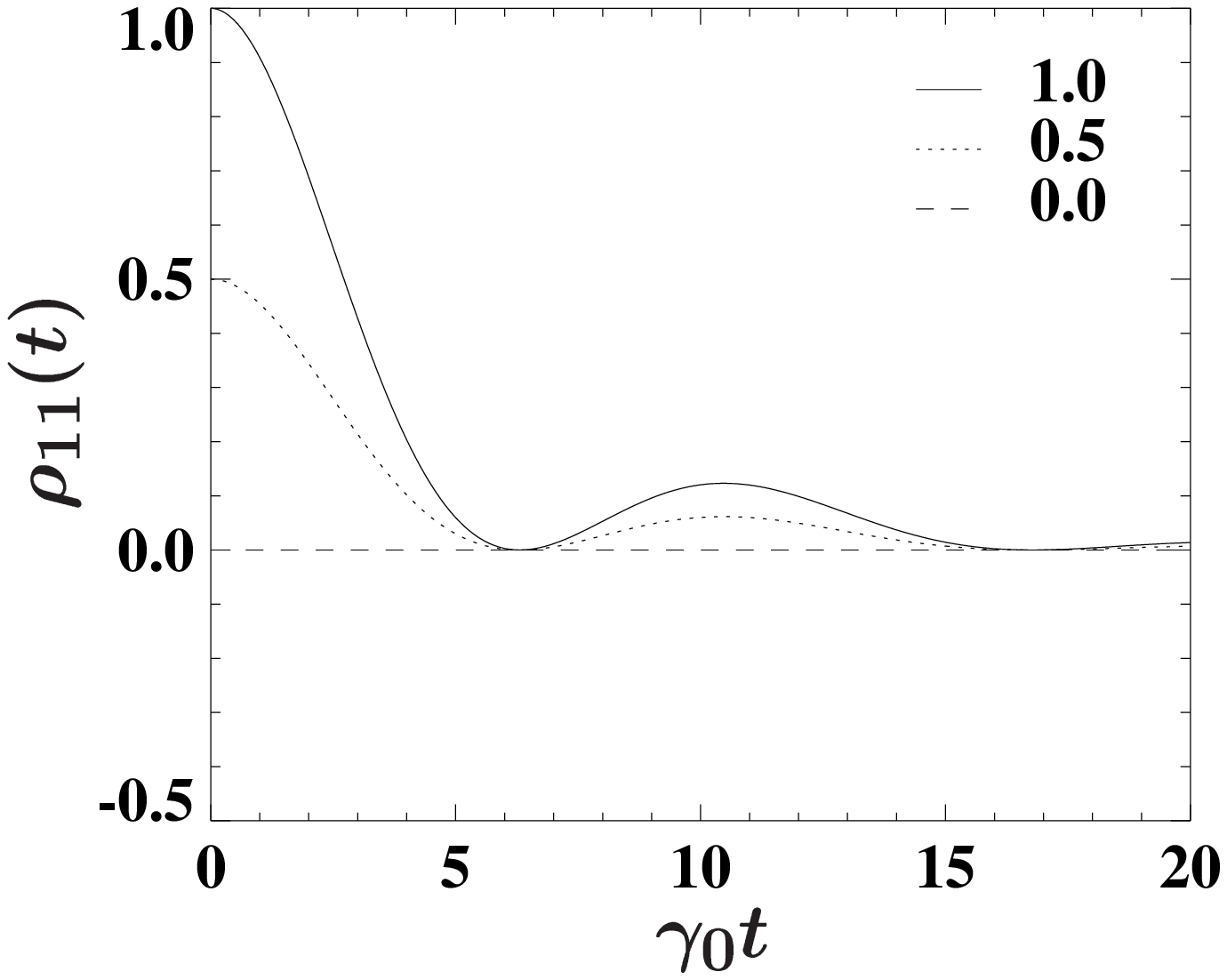}
    \caption{The damped Jaynes-Cummings model on resonance. Exact
      population for the three different initial conditions
      $\rho_{11}(0)=1.0$, $0.5$, $0.0$ in the strong coupling regime
      ($\tau_{\rm S}=0.2\tau_{\rm R}$).}
  \label{fig:2}
  \end{center}
\end{figure}

The 'failure' of the time-convolutionless master equation at $t=t_0$
can also be understood from a more intuitive point of view. The
time-convolutionless equation of motion (\ref{eq:tcl_ex}) states that
the time-evolution of $\rho(t)$ only depends on the actual value of
$\rho(t)$ and on the generator $K(t)$. However, at $t=t_0$ the time
evolution also depends on $\rho(0)$. This fact can be seen in
Fig.~\ref{fig:2}, where we have plotted $\rho_{11}(t)$ for three
different initial conditions, namely $\rho_{11}(0)=1.0$, $0.5$,
$0.0$.  At $t=t_0$, the corresponding density matrices coincide, 
regardless of the initial condition. However, the future time evolution 
for $t>t_0$ is different for these trajectories. It is therefore 
intuitively clear that a time-convolutionless form of the equation of 
motion which is local in time ceases to exist for $t>t_0$. 
The formal reason for this fact is that at $t=t_0$ the operator 
$1-\Sigma(t)$ (see Sec.~\ref{sec:tcl}), is not invertible and hence the 
generator $K(t)$ does not exist at this point.

\subsection{The damped Jaynes-Cummings model with detuning}
\label{sec:dJCdet}

\begin{figure}[t]
  \begin{center}
    \leavevmode \epsfxsize.49\linewidth\epsffile{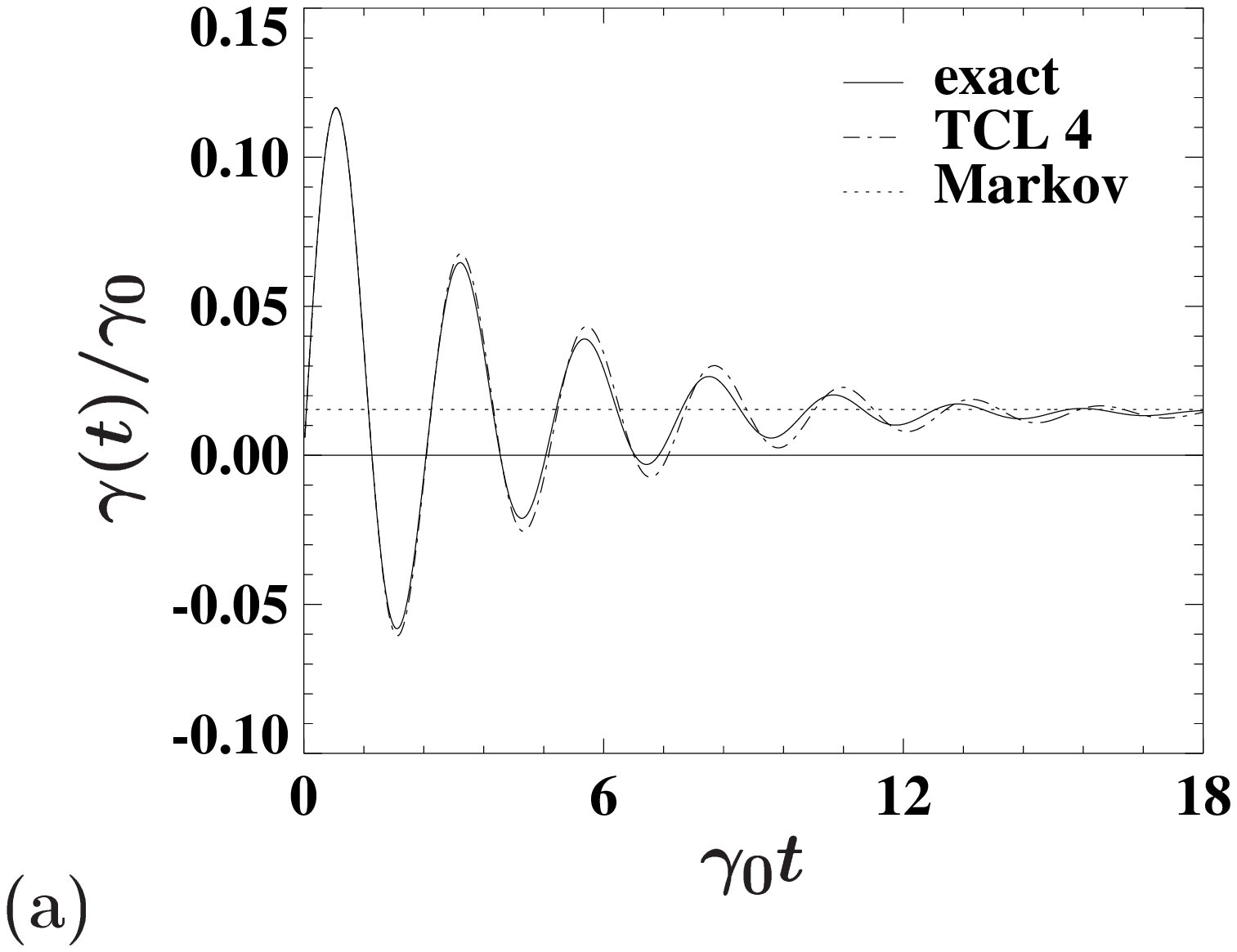}
    \leavevmode \epsfxsize.49\linewidth\epsffile{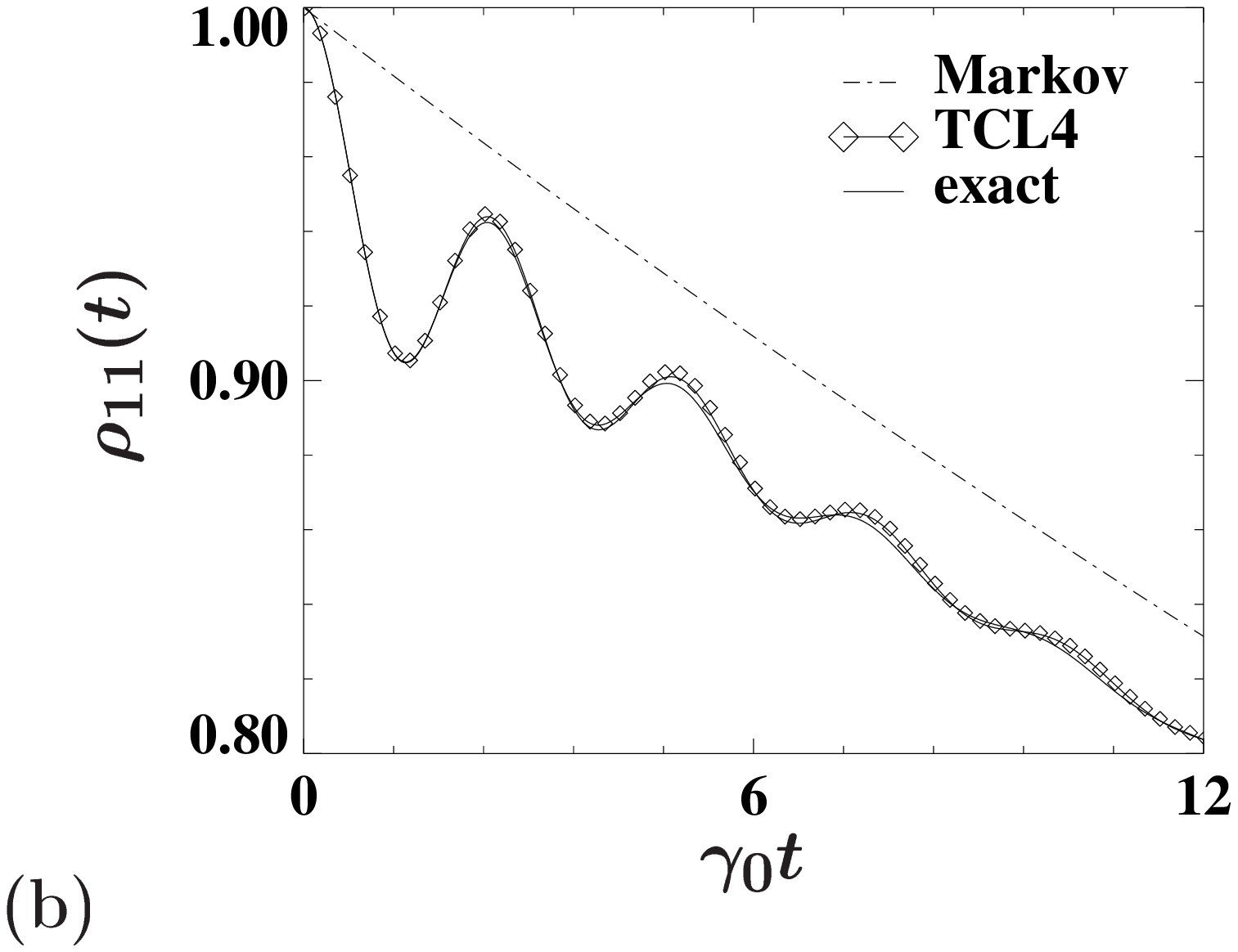}
    \caption{The damped Jaynes-Cummings model with detuning. Exact
      solution (exact), time-convolutionless master equation to fourth
      order (TCL 4), and the Markovian quantum master equation
      (Markovian): (a) Decay rate of the excited state population, and
      (b) the population of the excited state, including a stochastic
      simulation of the time-convolutionless quantum master equation
      with $10^5$ realizations, for $\lambda=0.3\gamma_0$ and $\Delta=8\lambda$}
  \label{fig:3}
  \end{center}
\end{figure}

In this section we treat the damped Jaynes-Cummings model with
detuning, i.~e., the same setup as in Sec.~\ref{sec:dJCres} but the
center frequency of the cavity $\omega_0$ is detuned by an amount
$\Delta=\omega_{\rm S}-\omega_0$ against the atomic transition
frequency. In this case the spectral density of the coupling strength reads
\begin{equation}
  \label{eq:JCdet}
  J(\omega)=\frac{1}{2\pi}\frac{\gamma_0\lambda^2} 
  {(\omega_{\rm S}-\Delta-\omega)^2+\lambda^2},
\end{equation}
and thus the functions $\Phi(t)$ and $\Psi(t)$ are given by 
\begin{eqnarray}
  \label{eq:phi}
  \Phi(t)&=&\gamma_0\lambda e^{-\lambda t}\cos(\Delta t),\\
  \label{eq:psi}
  \Psi(t)&=&\gamma_0\lambda e^{-\lambda t}\sin(\Delta t).
\end{eqnarray}
With these functions, the time-dependent Lamb shift and decay rate to
fourth order in the coupling, $S^{(4)}(t)$ and $\gamma^{(4)}(t)$,
respectively, can be calculated using Eqs.~(\ref{eq:Lamb}) and
(\ref{eq:rate}). The integrals can be evaluated exactly and lead to the
expressions
\begin{eqnarray}
  \label{eq:lamb_det}
  \lefteqn{S^{(4)}(t)=\frac {\gamma_0\lambda\Delta}{\lambda^2+\Delta^2}
   \left[1-e^{-\lambda t}\left(\cos(\Delta t)
  +{\textstyle\frac{\lambda}{\Delta}}\sin(\Delta t)\right)\right]}\nonumber\\
 && -\frac{\gamma_0^2\lambda^2\Delta^3e^{-\lambda t}}{2(\lambda^2+\Delta^2)^3}\Big\{
   \Big[1-3\left({\textstyle\frac{\lambda}{\Delta}}\right)^2\Big]\left(e^{\lambda
  t}-e^{-\lambda t}\cos(2\Delta t)\right)\nonumber\\
  && -2\Big[1-\left({\textstyle\frac{\lambda}{\Delta}}\right)^4\Big]\Delta
  t\sin(\Delta t)
   +4\Big[1+\left({\textstyle\frac{\lambda}{\Delta}}\right)^2\Big]\lambda
  t\cos(\Delta t)\nonumber\\
  &&-{\textstyle\frac{\lambda}{\Delta}}\Big[3-\left({\textstyle\frac{\lambda}{\Delta}}\right)^2\Big] 
  e^{-\lambda t}\sin(2\Delta t) 
  \Big\}
\end{eqnarray}
and
\begin{eqnarray}
  \label{eq:rate_det}
  \lefteqn{\gamma^{(4)}(t)=\frac {\gamma_0\lambda^2 }{\lambda ^{2}+\Delta ^{2}}
  \left[1-{e^{-\lambda t}}\left(\cos(\Delta t)
  -{\textstyle\frac{\Delta}{\lambda}}\sin(\Delta t)\right)\right]}\nonumber\\
 && +\frac{\gamma_0^2\lambda^5e^{-\lambda t}}{2(\lambda^2+\Delta^2)^3}\Big\{
   \Big[1-3\left({\textstyle\frac{\Delta}{\lambda}}\right)^2\Big]\left(e^{\lambda
  t}-e^{-\lambda t}\cos(2\Delta t)\right)\nonumber\\
  && -2\Big[1-\left({\textstyle\frac{\Delta}{\lambda}}\right)^4\Big]\lambda
  t\cos(\Delta t)
   +4\Big[1+\left({\textstyle\frac{\Delta}{\lambda}}\right)^2\Big]\Delta
  t\sin(\Delta t)\nonumber\\
  &&+{\textstyle\frac{\Delta}{\lambda}}\Big[3-\left({\textstyle\frac{\Delta}{\lambda}}\right)^2\Big] 
  e^{-\lambda t}\sin(2\Delta t) 
  \Big\}.
\end{eqnarray}
In Fig.~\ref{fig:3} (a) we have depicted $\gamma^{(4)}(t)$ together
with the exact decay rate, which can be calculated using the methods
of poles \cite{GarrawayPRA55} for $\Delta=8\lambda$ and
$\lambda=0.3\gamma_0$. Note, that the spontaneous decay rate is
severely suppressed compared to the spontaneous decay  on resonance.
This can also be seen by computing the Markovian decay rate
$\gamma_{\rm M}$ which is given by
\begin{eqnarray}
  \label{eq:Mark_det}
  \gamma_{\rm M}=\frac{\gamma_0\lambda^2}{\lambda^2+\Delta^2}\approx
  0.015\gamma_0.
\end{eqnarray}
However, this strong suppression is most effective in the long time
limit. For short times, $\gamma(t)$ oscillates with a large amplitude
and can even take negative values, which leads to an increasing
population. This is due to photons which have been emitted by the atom
and reabsorbed at a later time. Hence, the exact quantum master
equation as well as its time-convolutionless approximation are not in
the Lindblad form (\ref{eq:QME}), but conserve the positivity of the
reduced density matrix! This is of course not a contradiction to the
Lindblad theorem, since a basic assumption of the Lindblad theorem is
that the reduced system dynamics constitutes a 1-parameter dynamical
semi-group. However, in our example this is not the case, since the
initial preparation singles out the specific time $t=0$ and the domain 
of the operator $K(t)$ is shrinking for increasing $t$.

Since the transition rate $\gamma^{(4)}(t)$ also takes negative
values, we can not use the stochastic simulation algorithm
presented in Sec.~\ref{sec:stlind} for a stochastic unravelling of the
time-convolutionless quantum master equation, but have to use the
simulation algorithm in the doubled Hilbert space (see
Sec.~\ref{sec:SGQME}). The dynamics of the stochastic wave function
$\theta(t)=(\phi(t),\psi(t))^T$, which is an element of the doubled
Hilbert space $\widetilde{\cal H}={\cal H}\oplus{\cal H}$ is governed
by the stochastic differential equation (\ref{eq:gen_un}), where the
operators $F$ and $J$ are given by
\begin{equation}
  \label{eq:F_dbl}
  F=-\frac{1}{2}\gamma^{(4)}(t)\left(\begin{array}{cc}
   \sigma^+\sigma^- &0\\
  0&\sigma^+\sigma^- 
  \end{array}\right)
\end{equation}
and
\begin{equation}
  \label{eq:J_dbl}
  J=\sqrt{|\gamma^{(4)}|}\left(\begin{array}{cc}
  \mbox{sign}\left(\gamma^{(4)}\right)\sigma^-&0\\
  0& \sigma^-
  \end{array}\right).
\end{equation}
The deterministic part of the time evolution is governed by the
nonlinear Schr\"odinger-type equation 
\begin{equation}
  \label{eq:G_det}
   \frac{\partial}{\partial t}\theta(t)=G(\theta,t)=i\left( F+\frac{1}{2}
  \frac{\left\| J\theta(t)\right\|^2}{\|\theta(t)\|^2}\right)
  \theta(t),
\end{equation}
which results in a continuous drift, whereas the jumps induce
instantaneous transitions of the form
\begin{equation}
  \label{eq:dbl_jump}
  \theta(t)\longrightarrow\frac{\|\theta(t)\|}{\|J\theta(t)\|}J\theta(t)
  \sim\left(
    \begin{array}{c}
      \mbox{sign}\left(\gamma^{(4)}\right)|0\rangle_{\rm S}\\
      |0\rangle_{\rm S}
    \end{array}
\right).
\end{equation}
If the rate $\gamma^{(4)}(t)$ is positive then this type of transition
leads to a positive contribution to the ground state population
$\rho_{00}(t)$, whereas a negative rate leads to a decrease of $\rho_{00}$.

In Fig.~\ref{fig:3} (b), we show the results of  a stochastic
simulation for $10^5$ realizations, together with the analytical solution of
the time-convolutionless quantum master equation and the exact
solution. Obviously, the agreement of all three curves is good
and the stochastic simulation algorithm works excellently even for
negative decay rates. In addition, we also show the solution of the
Markovian quantum master equation which clearly underestimates the
decay for short times and does not show oscillations.

\subsection{Spontaneous decay into a photonic band gap}
\label{sec:gap}

\begin{figure}[t]
  \begin{center}
    \leavevmode \epsfxsize.49\linewidth\epsffile{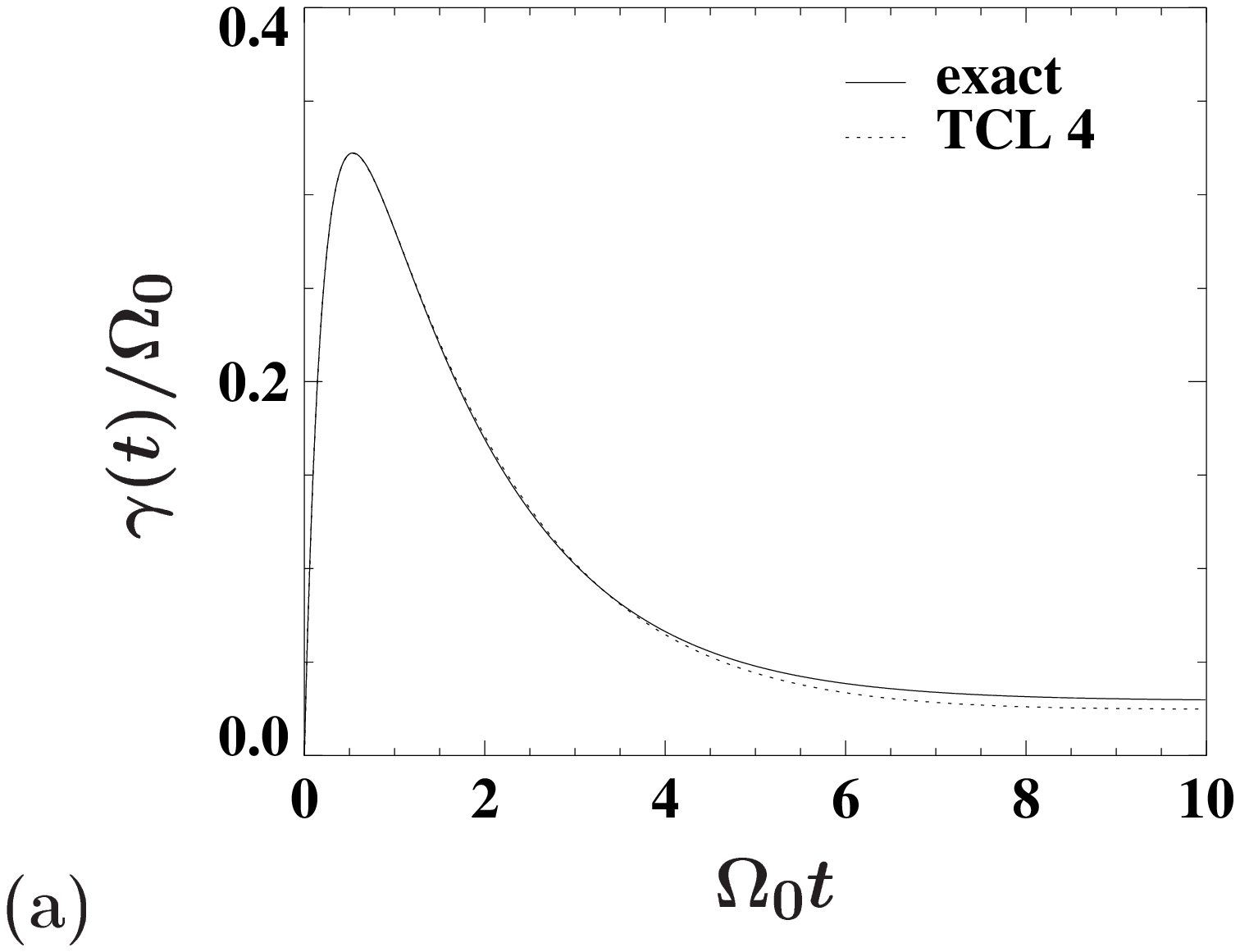}
    \leavevmode \epsfxsize.49\linewidth\epsffile{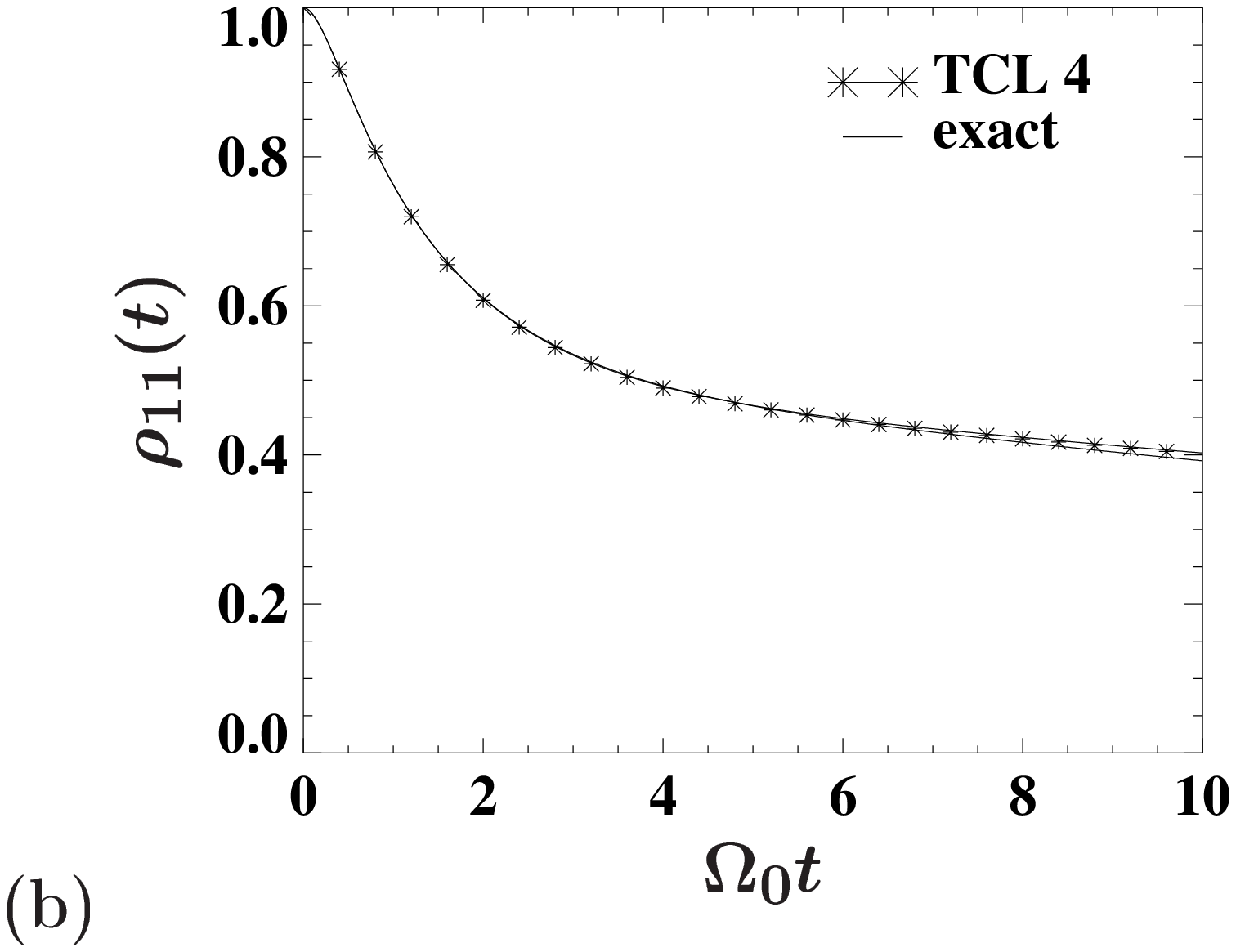}
    \caption{Spontaneous decay in a photonic band gap: Exact
      solution (exact), and time-convolutionless master equation to
      fourth order (TCL 4): (a) Decay rate of the excited state
      population, and (b) the population of the excited state,
      including a stochastic simulation of the time-convolutionless
      quantum master equation with $10^5$ realizations, for $W_1=1.1$,
      $W_2=0.1$, $\Gamma_1/\Omega_0=10$, and $\Gamma_2/\Omega_0=1$. }
  \label{fig:4}
  \end{center}
\end{figure}

As our final example, we treat a simple model for the
spontaneous decay of a two-level system in a photonic band gap which
was introduced by Garraway \cite{GarrawayGap}. To this end, we
consider a spectral density of the coupling strength of the form
\begin{eqnarray}
  \label{eq:J_gap}
  J(\omega)&=&\frac{\Omega_0^2}{2\pi}
 \bigg(\frac{W_1\Gamma_1}{(\omega-\omega_{\rm S})^2 +(\Gamma_1/2)^2}\nonumber\\
  &&\hspace*{2em}-\frac{W_2\Gamma_2}{(\omega-\omega_{\rm S})^2 +(\Gamma_2/2)^2}\bigg),
\end{eqnarray}
where $\Omega_0^2$ describes the overall coupling strength, $\Gamma_1$
the bandwidth of the 'flat' background continuum, $\Gamma_2$ the
width of the gap, and $W_1$ and $W_2$ the relative strength of the
background and the gap, respectively. Again, the function $J(\omega)$
has a small number of poles, and hence the exact solution can be
determined by using pseudomodes \cite{GarrawayGap}.  In
Fig.~\ref{fig:4} (a) we show the excited state's decay rate
$\gamma(t)$ for the same parameters as in Ref.~\cite{GarrawayGap},
i.~e., $\Gamma_1/\Omega_0=10$, $\Gamma_2/\Omega_0=1$, $W_1=1.1$, and
$W_2=0.1$. For short times, $\gamma(t)$ increases linearly on a
time scale of $\Gamma_1^{-1}$ and then takes a maximum, which stems
from transitions into the 'flat' background continuum. For longer
times, i.e., $t\gg\Gamma_2^{-1}$, the transitions into the background
are suppressed, and the decay rate becomes smaller and smaller until
it reaches its final value. Thus, the population of the excited state
decreases rapidly for times of the order $\Gamma_2^{-1}$, and slowly
in the long-time limit (see Fig.~\ref{fig:4} (b)).

The time-dependent Lamb shift $S^{(4)}(t)$ and the decay rate
$\gamma^{(4)}(t)$ of the time-convolutionless quantum master equation
to fourth order can be computed by inserting the spectral density of
the coupling strength $J(\omega)$ into Eq.~(\ref{eq:phi_psi}).  This yields
$\Psi(t)\equiv 0$ and
\begin{equation}
  \label{eq:Phi_gap}
  \Phi(t)=2\Omega_0^2\left(W_1 e^{-\Gamma_1 t/2}-W_2 e^{-\Gamma_2 t/2}\right),
\end{equation}
which can be inserted into Eqs.~(\ref{eq:Lamb}) and
(\ref{eq:rate}). Since $\Psi(t)\equiv 0$ the Lamb shift $S^{(4)}(t)$
vanishes; the time-dependent decay rate $\gamma^{(4)}(t)$ can be
computed explicitly, and is in good agreement with the exact decay
rate for our choice of parameters (see Fig.~\ref{fig:4} (a) and (b)).

\section{Summary}
\label{sec:con}

In this article we have presented a generalization of the stochastic wave
function method to quantum master equations which are not in
Lindblad form. This generalization -- together with the use of the
time-convolutionless projection operator technique -- makes it
possible to extend the range of potential applications of the
stochastic wave functions method beyond the weak coupling regime,
where the Born-Markov approximation is valid, without enlarging the
system. This generalization is capable of treating systems in the
intermediate coupling regime, i.~e., systems for which the generator
of the time-convolutionless quantum master equation exists for all $t$
and is analytic in the coupling strength $\alpha$. In the examples we
investigated in this article, this range was limited by $\tau_{\rm
S}\gtrsim \tau_{\rm R}$. The dynamics of this class of systems is
governed by an equation of motion which is local in time and can be
approximated by a perturbative expansion. This perturbative expansion
leads in general to a quantum master equation, which needs not to be
in Lindblad form but can be unravelled with our method. The basic idea
of this unravelling is the introduction of stochastic processes in a
doubled Hilbert space, which has been already successfully used for
the computation of matrix elements of operators in the Heisenberg
picture and multitime correlation functions. 

\vspace{3mm}

\noindent
{\bf{Acknowledgment}} \\
HPB would like to thank the Istituto Italiano per gli Studi Filosofici
in Naples (Italy) and BK would like to thank the DFG-Graduiertenkolleg
{\it Nichtlineare Differentialgleichungen} at the
Albert-Ludwigs-Universit\"at Freiburg for financial support of the
research project.


\bibliographystyle{prsty}  

\newpage
\end{multicols}
\widetext
\newpage

\end{document}